\documentclass[aps,prl,showpacs,twocolumn,groupedaddress,amssymb]{revtex4}

\usepackage{graphicx}
\usepackage{dcolumn}
\usepackage{bm}

\begin{document}


\title{Overcoming the Child Langmuir law via Pondermotive Potential of the intense lasers}

\author{S. Son}
\affiliation{18 Caleb Lane, Princeton, NJ 08540}
\author{Sung Joon Moon}
\affiliation{PACM, Princeton University, Princeton, NJ 08544}
\date{\today}

\begin{abstract}
The maximum current that could be carried in a vacuum tube
is given by the so-called Child Langmuir law.
A scheme where this limit could be overcome via the pondermotive potential is proposed.  
The estimation shows that the limit could be  overcome by a couple of hundred percents. 
\end{abstract}

\pacs{41.20.-q,  52.38.-r,  42.60.-v}                   
\maketitle

The current in the vacuum tube is limited by the space charge effect as given in the Child Langmuir law~\cite{child, qchild, qchild2}.  
As the intense current builds up inside the tube, 
the space-charge effects require more curvature 
in the static potential and at certain level of the current, 
the potential cannot maintain the desirable feature that 
the values of the static potential between the cathode and the anode
 are always larger than the value of the potential at the cathode.
Let us call this feature as the non-negative feature of the potential 
in the vacuum tube,  
The maximum current at which this feature is maintained is 
given by the so-called Child Langmuir law. 
The primary reason behind the requirement for the non-negative feature  
is due to the fact that 
the electrons, which are emitted at the cathode with the zero kinetic energy, 
should be able to reach the anode. 
In this paper, the pondermotive potential of the intense laser is contemplated in order to break the constraint. 
In other words, we consider the situation when the electrons can reach the anode even if the static potential does not satisfy the non-negative feature. 
We estimate  the physical parameter regime  which might be feasible 
for the practical use.
\begin{figure}
\scalebox{1.2}{
\includegraphics{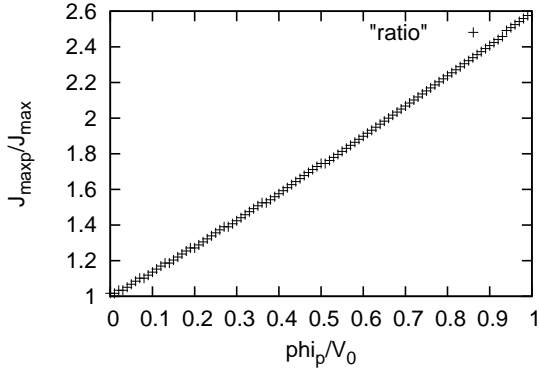}}
\caption{\label{fig:1}
The ratio $J_{\mathrm{maxp}}/ J_{\mathrm{max}}$ as a function of $\phi_p / V_0$ 
when  $ 2k = 1/d$. 
}
\end{figure}

To begin with, let us write down the 1-D fluid equations of the electrons in a vacuum tube.  The continuity equation, the momentum equation and the Poisson equation are given as 
\begin{eqnarray} 
  \frac{ \partial n_e(x) }{ \partial t }  + \frac{\partial (n_e v_x)}{ \partial x} = 0   \mathrm{,}\nonumber \\ \nonumber \\
\frac{ \partial v_x(x) }{ \partial t }  +  v_z\frac{\partial v_z}{ \partial x} 
= \frac{e}{m_e} \frac{ \partial \phi }{ \partial x } \mathrm{,} \nonumber \\ \nonumber \\
\frac{ \partial^2 \phi(x) }{ \partial x^2 } = 4\pi n_e(x) e \nonumber \mathrm{,} \\ \nonumber  
\end{eqnarray}
where $n_e(x)$ is the electron density between the cathode and the anode, 
 $v_x(x)$ is the electron velocity, and $\phi(x) $ is the static potential.  
The boundary condition would be $ v_x(0) = 0 $, $\phi(0) = 0 $ 
and $\phi(d) = V_0 $, 
where the distance between the cathode and the anode is $d$, 
and the potential has the bias of $V_0$ between the cathode and the anode. 
In the steady state, the above equation is simplified to 
$n_e(x) v_x(x) = \mathrm{const} $ and $(1/2 m_e v_x(x)^2 - e\phi(x)) = \mathrm{const}$ along with the same Poisson equation. 
Then, the solution can be simplified to 

\begin{equation} 
 \frac{ \partial^2 \phi(x) }{ \partial x^2  } =
\frac{4\pi J}{ \sqrt{ \frac{2e}{m_e} \phi(x)}} \mathrm{,}
\label{eq:sta}
\end{equation}
where $J = n e v_x(x) $ is the current density. 
The maximum current, at which the non-negative feature    can be maintained,
can be obtained trivially to be given as 
\begin{equation}
  J_{\mathrm{max}} = \frac{4}{9} \left( \frac{2e}{m_e}\right) \frac{1}{ 4 \pi} 
  \frac{V_0^{3/2}}{d^2} \mathrm{.}
\end{equation}
This is the Child Langmuir law. 

Now, consider a case when two-counter propagating lasers with the same frequency
are added in the vacuum tube.
The pondermotive potential is given as 
\begin{eqnarray} 
e\phi_p  =  \frac{m_e}{2} \frac{eE_1}{m_e \omega} \frac{e E_2 }{m_e \omega}
\cos(k x - \omega +\phi_1) \cos(kx+\omega+\phi_2)  \nonumber \\ \nonumber \\
\cong     \frac{m_e}{2} \frac{eE_1}{m_e \omega} \frac{e E_2 }{m_e \omega} \cos(2 kx + \phi_0)\mathrm{,} \nonumber \\ \nonumber 
\end{eqnarray} 
where $\phi_{ 1, 2} $ is the laser phase, $\phi_0 = \phi_1 - \phi_2$,  $\omega$ ($k$) is the laser frequency (wave vector), 
 and $E_1$ ($E_2$) is the electric field of each laser.  
The two lasers have the same frequency so that 
the pondermotive potential is time-independent. 
Then, there  is a steady state solution.  
The continuity equation is still given as  $n_e(x) v_x(x) = \mathrm{const} $
Due to the pondermotive potential, 
the momentum  and Poisson equation is modified as

\begin{eqnarray} 
  \frac{1}{2} m_e v_x(x)^2 - e(\phi(x) +\phi_p(x)) = \mathrm{const}  \mathrm{,} 
\nonumber \\ \nonumber \\ 
 \frac{ \partial^2 \phi(x) }{ \partial x^2  } =
\frac{4\pi J}{ \sqrt{ \frac{2e}{m_e} (\phi(x)+\phi_p(x))}} \mathrm{,} \nonumber \label{eq:sta3} \nonumber \\
 \nonumber 
\end{eqnarray}
The  non-negative feature   can be recasted in the new setting as $ \phi(x) + \phi_p(x) > \phi(0) + \phi_p(0) $ for $0<x<d$.  If we choose the phase $\phi_0$  so that $\phi_p(0)=0$,  then the condition for  the non-negative feature   is 
\begin{equation}
\phi(x) + \phi_p(x) > 0 \mathrm{.}
\end{equation}
Noting that the condition for  the non-negative feature  
without the pondermotive potential is $\phi(x) > 0 $,  
it  is  possible to flow the electrons from the cathode to the anode 
even though $\phi(x) <0 $ at some point because the potentials need to satisfy only the condition, $\phi(x) + \phi_p(x) > 0$. 
The maximum current $J_{\mathrm{maxp}} $ can be determined trivially from Eq.~(\ref{eq:sta3}).
In Fig.~(\ref{fig:1}), we plot the $J_{\mathrm{maxp}}/ J_{\mathrm{max}}$ is plotted as a function of $\phi_p / V_0$ when $ 2k = 1/d$.  As shown, the ratio is an increasing function of $\phi_p / V_0$.  

Now, the estimation of practically
 meaningful  parameter regimes is in order. 
The powerful vacuum device has the dimension of  1 cm to 10 cm. 
Since the wave vector of the pondermotive potential 
should be comparable to the dimension of the devices. 
The wave length of the laser should be order  of between 1 cm to 10 cm, 
which corresponds to 3 to 30 GHz of the frequency.
Another constraint would be that the strength of the pondermotive potential should be comparable to the bias between the cathode and the anode. 
Assuming the same pair of the lasers, 
the pondermotive potential could be estimated as
\begin{equation}
\phi_p \cong 2 \times 10^{-5} \lambda^2 I \mathrm{,}
\end{equation}
where $I$ is the intensity of the laser in the unit of $ W/\mathrm{cm}^2 \sec $,
$\lambda$ is the wave length of the laser in the unit of $\mathrm{cm} $. 
Given the distance between the anode and the cathode is $d$, let us denote the 
the cross-section of the anode (cathode) as $s^2 = \alpha^2 d^2$.
Let us also denote the wave length of the laser as $ d= \beta \lambda$.  
Then, the minimum total power of the lasers  would be  given as 
$P =   I s^2 = (0.5 \times 10^4 / \lambda^2) s^2 \phi_P$, 
where $\phi_p $ is in the unit of volt.   
By assuming $\phi_p = \gamma V_0$, 
the total power of laser can be estimated as 
 \begin{equation} 
P =  0.5 \times 10^4 \alpha^2 \beta^2 \gamma  V_0 \mathrm{.} 
\end{equation} 
For $\alpha =1 $,$\beta =1$,  $\gamma =1 $ and  $V_0 =  1  \ \mathrm{kV}$,  
the laser power should be 50 MW.  
However, if $\alpha = 0.3$,  $\beta = 0.3$ and $\gamma=0.1$, 
 the laser power requirement would be 500 kW for $V_0 = 1 \ \mathrm{kV}$ and 
 50 kW for  $V_0 = 100 \ \mathrm{V}$.  
These powers might be delivered using the powerful mirco-wave devices such as the gyrotron and magnetron. 

In summary, we have proposed a way to overcome the Child Langmuir wave via the pondermotive potential of the intense laser.  The regime of the practical use has been estimated. 
While we uses two identical lasers in order to achieve the goal,  the laser with the different frequency can be also used  in the context of the Raman scattering~\cite{sonlandau, sonbackward, sonprl, sonpla, Fisch, Fisch3}.  
In comparison to the scheme discussed here, the Raman scattering would produce much more intense Langmuir wave from the pondermotive potential
 so that the effect by the lasers would be enhanced. On the other hand,
 the time dependency of the Langmuir wave complicate the matter so that the reformulation of the Child Langmuir law in the time dependent setting should be worked out~\cite{fchld}.  
This would be an interesting result topic. 
While the quantum version of the Child Langmuir law~\cite{qchld, qchild, qchild2} 
has been devised,  
the discussion in this paper 
would be relevant due to the fact that 
the hard x-ray laser would be available 
due to the free electron laser~\cite{Free2}.  
The analysis of the quantum Child Langmuir law 
in the context of the x-ray pondermotive force 
could be also an interesting research topic.


\end{document}